\documentclass[conference,twocolumn]{IEEEtran}

\usepackage[T1]{fontenc}

\usepackage{cite}

\makeatletter
\newcommand{\distas}[1]{\mathbin{\overset{#1}{\kern\z@\sim}}}%
\newsavebox{\mybox}\newsavebox{\mysim}
\newcommand{\distras}[1]{%
  \savebox{\mybox}{\hbox{\kern3pt$\scriptstyle#1$\kern3pt}}%
  \savebox{\mysim}{\hbox{$\sim$}}%
  \mathbin{\overset{#1}{\kern\z@\resizebox{\wd\mybox}{\ht\mysim}{$\sim$}}}%
}
\makeatother

\ifCLASSINFOpdf
  \usepackage[pdftex]{graphicx}
\else
  \usepackage[dvips]{graphicx}
\fi
\usepackage{caption}
\usepackage{amsmath}

\usepackage[cmintegrals]{newtxmath}

\usepackage{algorithmic}

\usepackage{array}

  \usepackage[caption=false,font=normalsize,labelfont=sf,textfont=sf]{subfig}

\usepackage{fixltx2e}

\begin{document}
%

\title{\huge{Energy-Aware Multi-Server Mobile Edge Computing: \\ A Deep Reinforcement Learning Approach}}

\author{
    \IEEEauthorblockN{Navid Naderializadeh\\
    Intel Corporation\\
    navid.naderializadeh@intel.com} \and
    \IEEEauthorblockN{Morteza Hashemi\\
    University of Kansas\\
    mhashemi@ku.edu}
}

\maketitle

\begin{abstract}
We investigate the problem of computation offloading in a mobile edge computing architecture, where multiple energy-constrained users compete to offload their computational tasks to multiple servers through a shared wireless medium. We propose a multi-agent deep reinforcement learning algorithm, where each server is equipped with an agent, observing the status of its associated users and selecting the best user for offloading at each step. We consider computation time (i.e., task completion time) and system lifetime as two key performance indicators, and we numerically demonstrate that our approach outperforms baseline algorithms in terms of the trade-off between computation time and system lifetime.


\end{abstract}


%
\IEEEpeerreviewmaketitle


\section{Introduction}
It is envisioned that 5G-and-beyond will enable an unprecedented proliferation of data-intensive and computationally-intensive applications such as face recognition, location-based augmented/virtual reality (AR/VR), and online 3D gaming~\cite{hashemi2018efficient,hashemi2017out,yao2019edgeflow, yang2018communication, 8319323}. However, adoption of these resource-hungry applications will be negatively affected by limited on-board computing and energy resources. In addition to the computationally-intensive applications, billions of IoT devices are expected to be deployed for various applications such as health monitoring, environmental monitoring and smart cities, to name a few. These applications require a large number of low-power and resource-constrained wireless nodes to collect, pre-process, and analyze huge amounts of sensory data~\cite{guo2018mobile}, which may not be feasible due to the limited on-board computing resources. 

\begin{figure}[t]
    \centering
    \includegraphics[scale=.33, trim = 2cm 2cm 2cm 1.5cm,clip]{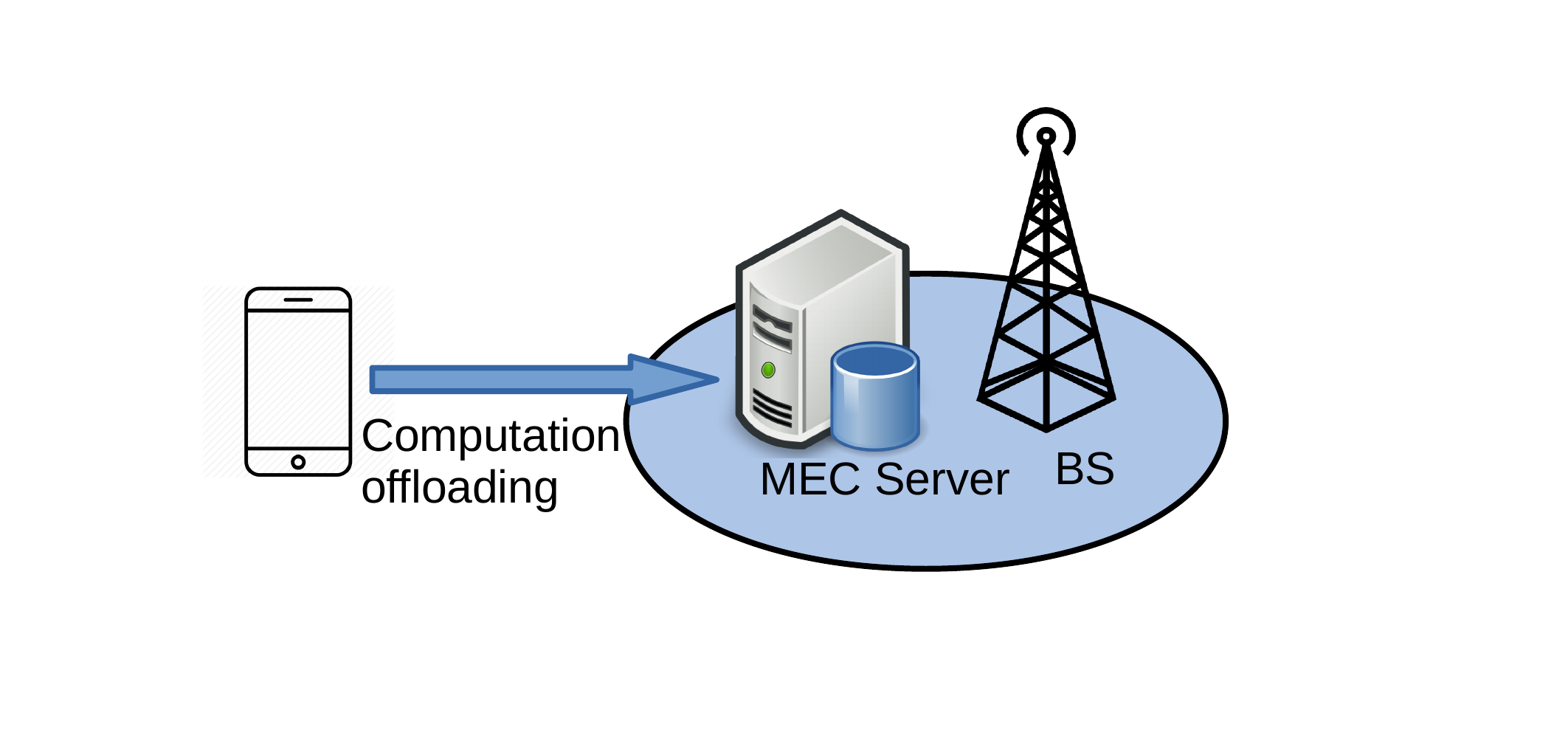}
    \caption{Mobile Edge Computing (MEC) system architecture.}
    \label{fig:MEC}
    \vspace{-.1in}
\end{figure}

In order to bridge the gap between increasing demand for mobile computational power and constrained on-board resources, mobile edge computing (MEC) has been contemplated as a solution to supplement the computing capabilities of the end-users~\cite{you2016multiuser, le2017efficient, messaoudi2017using, li2018incentive, chen2018decentralized}. In contrast to the traditional cloud computing architectures, such as Amazon Web Services (AWS) and Microsoft Azure, MEC leverages the radio access networks to boost the computing power in close proximity to end-users, thus enabling users to offload their computations to MEC servers, as shown in Figure~\ref{fig:MEC}.

Under the MEC model, each user either offloads its computation to the server or uses its own resources to locally perform the computation. In this case, users can save energy and prolong the overall lifetime of the system by offloading to the central node (assuming central node is not energy sensitive). However, if all users offload their computations to the central node, on one hand the communication resources need to be divided among all users, which decreases the effective uplink throughput, and on the other hand, the queuing delay and computation time at the central node increases. Therefore, a dynamic policy to select the ``best'' offloading user is needed in order to strike the optimal trade-off between the lifetime of the system and the computation time. Thus, we note that before a practical MEC architecture becomes a reality, it faces several challenges including \emph{efficient management of communication and computing resources and coordination among distributed users and several base stations.} 

In practical MEC scenarios, \emph{the system is partially observable} in the sense that users are distributed  and the central node only observes the state (e.g., energy level and computation load) of those users that have offloaded so far. In addition, imperfect and delayed channel state information (CSI) makes the problem even more challenging since the central node needs to optimally balance the intricate ``exploration and exploitation'' trade-offs, i.e., to exploit those users with more up-to-date information or to explore those users which have not offloaded yet or their state information is not fresh. 

In this paper, we consider a MEC architecture involving multiple users and multiple MEC servers. 
The reason we focus on a multi-server architecture is due to the fact that densification of small cells with abundant amounts of computational power is a key technique for improving the system throughput in 5G networks and beyond~\cite{naderializadeh2017ultra, naderializadeh2018feedback}. In such a scenario, we develop an autonomous and energy-aware distributed computing platform via multi-agent deep reinforcement learning, whose objective is to increase the lifetime of the system, as well as to decrease the average duration for computing incoming tasks to the users. We show, through simulation results, that our proposed approach strikes the right trade-off between the aforementioned metrics, outperforming two greedy baseline algorithms.


\section{Background}\label{sec:background}
\subsection{Reinforcement Learning}
Reinforcement learning (RL) is a type of machine learning, in which an agent or a group of agents interact(s) with an environment by collecting \textbf{observations}, taking \textbf{actions}, and receiving \textbf{rewards}.
The agent's experience is given by the tuple $(s_t, a_t, r(s_t,a_t), s_{t+1})$ such that at time step $t$, the agent observes current state of the environment denoted by $s_t$, and chooses action $a_t$, which results in a reward $r(s_t,a_t)$. The state will transition to $s_{t+1}$ according to the transition probability $p(s_{t+1}|s_t, a_t).$ The ultimate goal for the agent is to \emph{learn} what action to take given each observation to maximize its cumulative reward over time.

Deep reinforcement learning has been proposed as an enhancement to more traditional RL approaches, where the agent uses a deep neural network as a function approximator to represent its policy and/or value function. This enables the observation space (and potentially the action space) to be continuous and uncountable. Deep Q-Network (DQN)~\cite{mnih2013playing} is a specific deep RL agent, where its state-action value function is updated by minimizing the following loss, which is derived through the \emph{Bellman Equation}: 
	\begin{align*}
	    L(\theta) = \mathbb{E}\left[ Q(s_t,a_t;\theta) - \left( r(s_t, a_t) + \gamma \max_{a_{t+1}} Q(s_{t+1},a_{t+1};\theta)\right) \right],
	\end{align*}
where $Q(s_t,a_t;\theta)$ represents the estimated state-action value function for state $s_t$ and action $a_t$ and the set of DQN parameters denoted by $\theta$. 

\begin{figure}[h]
\vspace{-.15in}
	    \centering
	    \includegraphics[scale=0.4, trim = 2cm 1.5cm 2cm 1.4cm, clip]{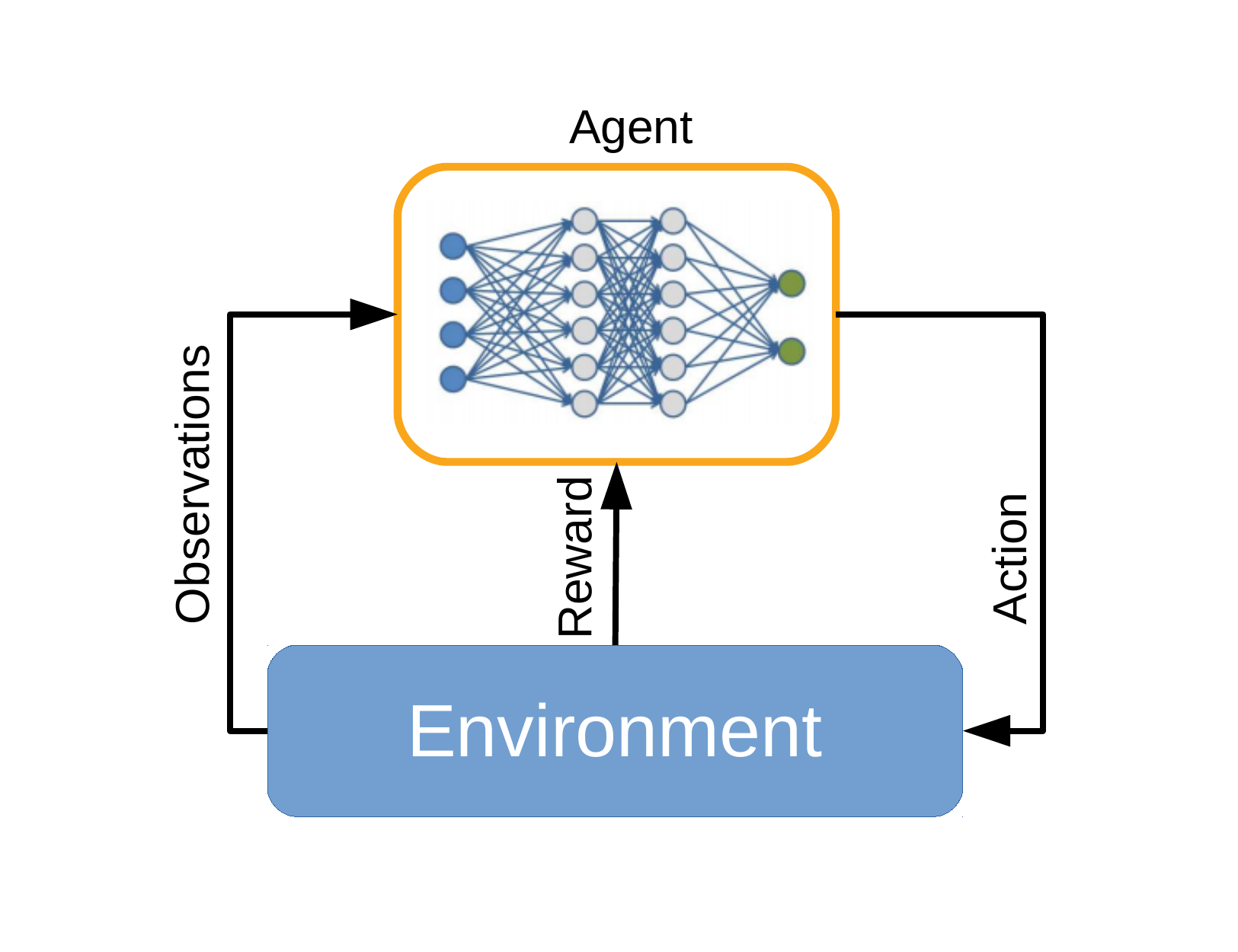}
	    \caption{Deep reinforcement learning model.}
	    \label{fig:DRL}
	    \vspace{-.15in}
	\end{figure}

\subsection{Related Work}
Recently, there has been an extensive amount of work investigating the mobile edge computing paradigm. Based on the number of users and servers, there are several architectures that have been investigated.

The work in \cite{yang2018communication} considers offloading with one base station (BS) and one user, where the user may offload a set of virtual reality (VR) tasks for computation at the BS, or it may compute them locally as well. An optimization problem is solved to schedule the tasks for computation at the user or the BS in order to minimize the average transmitted data per task. Similarly, the authors in \cite{messaoudi2017using} consider a single-user scenario with multiple tasks, some of which can be offloaded to a central server. The tasks have dependency, which is represented by a graph. The graph is partitioned to multiple clusters, and then an integer programming problem is formulated to determine whether to offload each cluster or not such that the total execution time of tasks is minimized given energy constraints.

The authors in \cite{you2016multiuser} consider a multi-user offloading problem with different computing tasks, each of which can be partially done by the user and the rest to be offloaded to a central server. The objective is to minimize the total weighted energy consumption (local computing plus offloading to central BS), subject to a total fixed delay constraint and computation capacity constraint at all users and central BS. In this case, weights are set arbitrarily, and users offload only using TDMA. In addition, communication time is ignored compared to computation time. The work in \cite{le2017efficient} considers TDMA and FDMA methods for the users to offload to the central server such that downlink delay is ignored. However, contrary to \cite{you2016multiuser}, the system considered in \cite{le2017efficient} does not necessarily assign computation resources proportionally to the size of offloaded tasks. The objective is to minimize the total computation delay (maximum of local computation delays and central computation plus offloading delays). 



The authors in \cite{chen2018decentralized} apply deep reinforcement learning to obtain efficient computation offloading policies independently at each mobile user. In this case, a continuous state space is defined and a deep deterministic policy gradient (DDPG) agent is adopted to handle the high-dimensional action space. Moreover, in \cite{mahn2018distributed}, an energy minimization offloading problem with a time constraint is tackled. A game theory approach is used to decompose the problem into two sub-problems such that first the access point (AP) receives the offloading decisions from users, and then it optimizes the communication and computation resources (e.g., channel access time and computation power allocated to each user). Based on the assigned resources, each user autonomously decides between local computation, offloading to AP, or offloading to the cloud. Users then report their decisions to the AP.  


\section{System Model}\label{sec:model}
We consider a network with $N$ MEC servers and $K$ users, which are located randomly within an $R\times R$ network area. The network operates in a time-slotted fashion, where the duration of each time interval is denoted by $\tau$. The users receive multiple computation tasks to complete over time. In order to do that, they have two options: i) compute the tasks locally, or ii) offload the tasks to be computed at one of the MEC servers.
We assume that all users start from a full energy level of $E_{\max}$, and then gradually consume energy over time until depletion, in which case the system's lifetime is over. We use $E_i(t)$ to denote the energy level of user $i$ at the beginning of time interval $t$.

\subsection{Task Arrival Process}
At time interval $t$, each individual user $j\in\{1,...,K\}$ receives a set of computation tasks, denoted by $\mathcal{T}_j(t)$ to compute. We assume a \textit{Poisson} arrival process for the tasks, where the number of incoming tasks at each time interval follows an i.i.d. Poisson distribution with rate $\lambda$; i.e.,
\begin{align}
|\mathcal{T}_j(t)| \distras{i.i.d.} \mathrm{Pois(\lambda)}, \forall j\in\{1,...,K\}, \forall t.
\end{align}
The tasks will be buffered in the user's queue and served on a first-in-first-out basis. We assume the tasks are homogeneous in size, implying that for any user $j\in\{1,...,K\}$ at any time interval $t$, every task in $\mathcal{T}_j(t)$ has a fixed size of $B$ bits.

\subsection{Local Computation Model}
As mentioned above, one way for each user to serve its incoming tasks is to compute the task using its local processor. We adopt a local computation model similar to~\cite{chen2018decentralized}, where the user first computes its maximum feasible computing power at any interval, and uses that to compute the maximum number of bits it can compute. To be precise, for user $j\in\{1,...,K\}$, the maximum feasible local computation power at time interval $t$ is calculated as:
\begin{align}\label{eq:power}
P_{j}^{\max}(t) = \frac{E_j(t)}{\tau}.
\end{align}
Then, the maximum feasible CPU frequency is computed as:
\begin{align}
f_{j,\mathrm{local}}^{\max}(t) = \min\left[f_{j,\mathrm{local}}^{\max}, \sqrt[3]{\frac{P_{j}^{\max}(t)}{\kappa}}\right],
\end{align}
where $f_{j,\mathrm{local}}^{\max}$ denotes the absolute maximum CPU frequency for user $j$, and $\kappa$ represents the effective switched capacitance. This will lead to the maximum number of bits that can be computed by user $j$ at time $t$ as
\begin{align}
B_{j,\mathrm{local}}^{\max}(t) = \left\lfloor \frac{\tau \times f_{j,\mathrm{local}}^{\max}(t)}{L_j} \right\rfloor,
\end{align}
where $L_j$ denotes the number of CPU cycles per bit at user $j$. The user then checks its task buffer, and computes the tasks at the head of the queue one by one as long as the total number of computed bits does not exceed $B_{j,\mathrm{local}}^{\max}(t)$. Note that if the size of the first task is already larger than $B_{j,\mathrm{local}}^{\max}(t)$, then the user remains idle and does not do any local computation at that step. We denote the effective consumed energy for the local computation of user $j$ at time interval $t$ by $E_{j,\mathrm{local}}(t)$.

\subsection{Task Offloading Model}
The other option for the users to compute their incoming tasks is to offload the tasks to the MEC servers. We assume that before the task arrival process begins, each user is associated with the MEC server which has the strongest long-term channel gain to it. We denote by $S_j$ the MEC server to whom user $j$ is associated, and by $\mathcal{U}_i$ the set of associated users to MEC server $i$. The local user pools of the MEC servers are disjoint; i.e., $\mathcal{U}_i \cap \mathcal{U}_{i'} = \emptyset, \forall i\neq i'$.

For user $j$ to offload its computation tasks to server $S_j$ at time interval $t$, it first calculates its maximum feasible transmit power based on its instantaneous energy level as in \eqref{eq:power}. It then obtains its maximum uplink achievable rate as
\begin{align*}
R_{j}^{\max}(t) = W_{S_j,j}(t) \log_2\left(1 + \gamma_{S_j,j}(t) \min\left[ P_{j}^{\max}(t), P_{j,\mathrm{Tx}}^{\max} \right] \right),
\end{align*}
where $W_{S_j,j}(t)$ denotes the amount of bandwidth allocated to the uplink transmission between user $j$ and server $S_j$ at time interval $t$, $\gamma_{S_j,j}(t)$ denotes the received signal-to-noise ratio (SNR) from user $j$ to server $S_j$ at time interval $t$, and $P_{j,\mathrm{Tx}}^{\max}$ denotes the absolute maximum transmit power of user $j$. The uplink transmissions of users to their respective MEC servers at each time interval may share the spectrum using multiple access techniques, such as FDMA or TDMA. Therefore, the maximum number of bits that user $j$ can transmit to server $S_j$ at time interval $t$ can be computed as
\begin{align}
B_{j,\mathrm{offload}}^{\max}(t) = \left\lfloor \tau \times R_{j}^{\max}(t)\right\rfloor.
\end{align}
Similar to local computation, the user offloads tasks from head of its task buffer whose total number of bits does not exceed $B_{j,\mathrm{offload}}^{\max}(t)$. We denote the effective consumed energy by user $j$ to offload its tasks to its associated server at time interval $t$ by $E_{j,\mathrm{offload}}(t)$.

\subsection{Energy Model}
We assume that at each interval, each user either stays idle, does local computation of tasks, or offloads some tasks to its serving MEC server. Denoting the action taken by user $j$ at time interval $t$ by $a_{j,t}$, the energy level of the user evolves over time as follows:
\begin{align*}
E_j(t+1) =
\begin{cases}
E_j(t) - \epsilon & \text{if }a_{j,t}= \text{idle,} \\
E_j(t) - E_{j,\mathrm{local}}(t) - \epsilon & \text{if }a_{j,t}= \text{local comp.} \\
E_j(t) - E_{j,\mathrm{offload}}(t) - \epsilon & \text{if }a_{j,t}= \text{offloading.}
\end{cases},
\end{align*}
where $\epsilon$ denotes the unit stand-by energy consumption for each user at every time interval.

\subsection{Problem Statement}
As mentioned before, we assume that the systemncrashes once at least one of the users runs out of energy. This leads to the definition of the \emph{system lifetime}, denoted by $\mathsf{LT}$, as follows:
\begin{align}
\mathsf{LT} = \max \left\{t \ | \ E_j(t) > 0, \forall j\in\{1,...,K\} \right\}.
\end{align}

Furthermore, for any incoming task $T$, let $b_T$ and $c_T$ respectively denote the time intervals when the task arrives and when the task computation is completed, either through local computation or offloading to the servers. We define the \emph{mean task completion time}, denoted by $\mathsf{TCT}$, as the average time it takes for a task to be computed before the system crashes; i.e.,
\begin{align}
\mathsf{TCT} = \frac{1}{|\mathcal{T}|} \sum_{T \in \mathcal{T}} c_T - b_T,
\end{align}
where $\mathcal{T}$ denotes the set of all completed tasks within the system lifetime, defined as:
\begin{align}
\mathcal{T} = \bigcup_{j=1}^K \bigcup_{t=1}^{\mathsf{LT}} \left\{T \in \mathcal{T}_j(t): c_T < \mathsf{LT} \right\}.
\end{align}

Having defined these metrics, our goal is to minimize the mean task completion time, while increasing the system lifetime as much as possible. Note that there is an inherent trade-off between these two metrics since reducing the mean task completion time requires more local computation and offloading to the MEC servers, which depletes the users' energy levels more quickly, hence reducing the system lifetime.

\section{Proposed Multi-Agent Deep Reinforcement Learning Approach}\label{sec:method}

\begin{figure}
    \centering
    \includegraphics[scale=.23, trim = 2cm 3.5cm 3cm 2cm, clip]{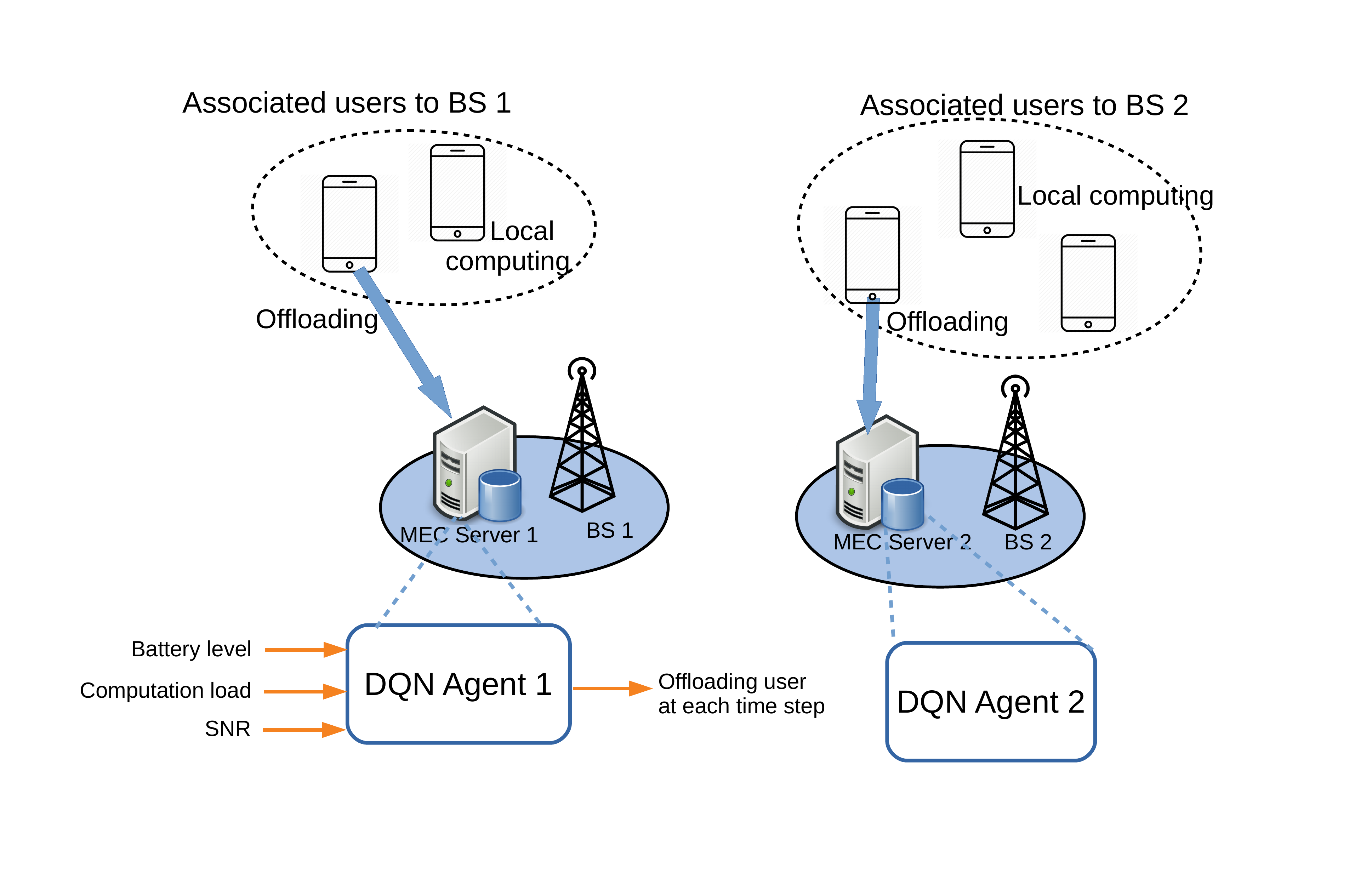}
    \caption{System model.}
    \label{fig:system_model}
    \vspace{-.2in}
\end{figure}

In order to enhance the trade-off between system lifetime and task completion time, we propose to equip each MEC server with a DQN agent, which  selects the best user (across its associated users) for offloading its tasks to the server at each time interval. The proposed model is shown in Figure~\ref{fig:system_model}.

In particular, we consider an episodic time frame, where at the beginning of each episode, the user and server nodes are dropped randomly within the network area, with user nodes at their maximum energy level. We then run the system until at least one of the nodes runs out of energy, in which case the episode terminates and the node locations, task buffers, and energy levels are reset for the next episode.


\subsection{Observations and Actions}
We assume that at the beginning of each time interval, the DQN agent at each MEC server receives a partial observation of the environment, including the queue length, energy level, mean task waiting time, and uplink SNR of its associated users, and then it decides which user from its local associated user pool should offload its tasks to the server. The rest of the users in the pool perform local computation of their tasks at that step provided that they have sufficient energy to do so.  

\subsection{Rewards}
As mentioned in Section~\ref{sec:model}, our ultimate goal is to increase the system lifetime and decrease the average time it takes to compute an incoming task. In order to do that, at each time interval, after the agents take their actions, we provide each agent with an individual reward
in the form of energy efficiency, i.e., the ratio of the selected user's computed bits (which were offloaded to the server) to the selected user's consumed energy for offloading.

\subsection{Numerical Results}\label{sec:sim}
We have conducted extensive simulations in order to evaluate the performance of our proposed approach. We consider a network area of size $10m \times 10m$. We assume the maximum energy level of each user at the beginning of each episode is selected uniformly at random from the interval $(0.01, 1)J$. The maximum transmit power of each user is taken to be $27$ dBm. We assume a time interval length of $\tau = 100 ms$. The server and user CPU frequencies are taken to be $3$ GHz and $1$ GHz, respectively, with respective cycles per bit of $1000$ and $500$. The effective switched capacitance is set to  $\kappa=10^{-27}$. The noise variance is taken to be $-174$ dBm/Hz, the total system bandwidth is set to $20$ MHz, and the transmissions are assumed to use FDMA. The mean task arrival rate is taken to be $10$, the task length is equal to $1$ KB and the unit stand-by energy is set to $\epsilon=10^{-7}$.

As for the DQN agent, we use a $2$-layer neural network with $200$ nodes per layer and tanh activation function. We use an $\epsilon$-greedy policy, with probability of random actions staying at $100\%$ for $100$ initial pre-training episodes, and then decaying to $1\%$ over $10^4$ time intervals. The experience buffer size is set to $10^5$ samples, and a discount factor of $\gamma=0.9$ is utilized. The agent is updated at the end of every episode, with a batch of size $64$ from the buffer. The learning rate also starts from $5\times10^{-3}$ and is cut in half every $100$ episodes.

\begin{figure*}[ht]
\centering
\subfloat[]{
  \includegraphics[trim = 0in 2.7in 0in 3in, clip,width=0.4\textwidth]{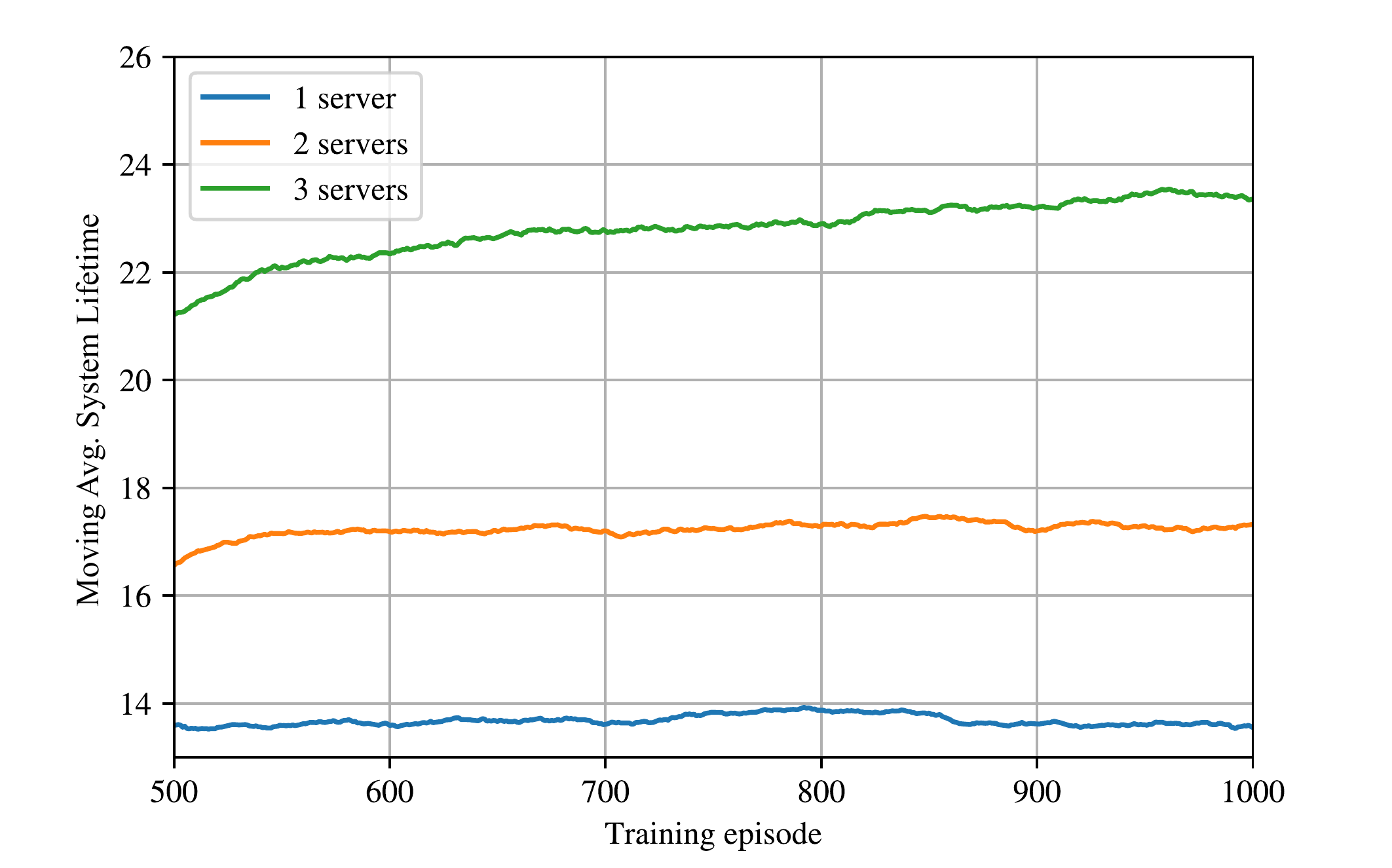}
}
\subfloat[]{
  \includegraphics[trim = 0in 2.7in 0in 3in, clip,width=0.4\textwidth]{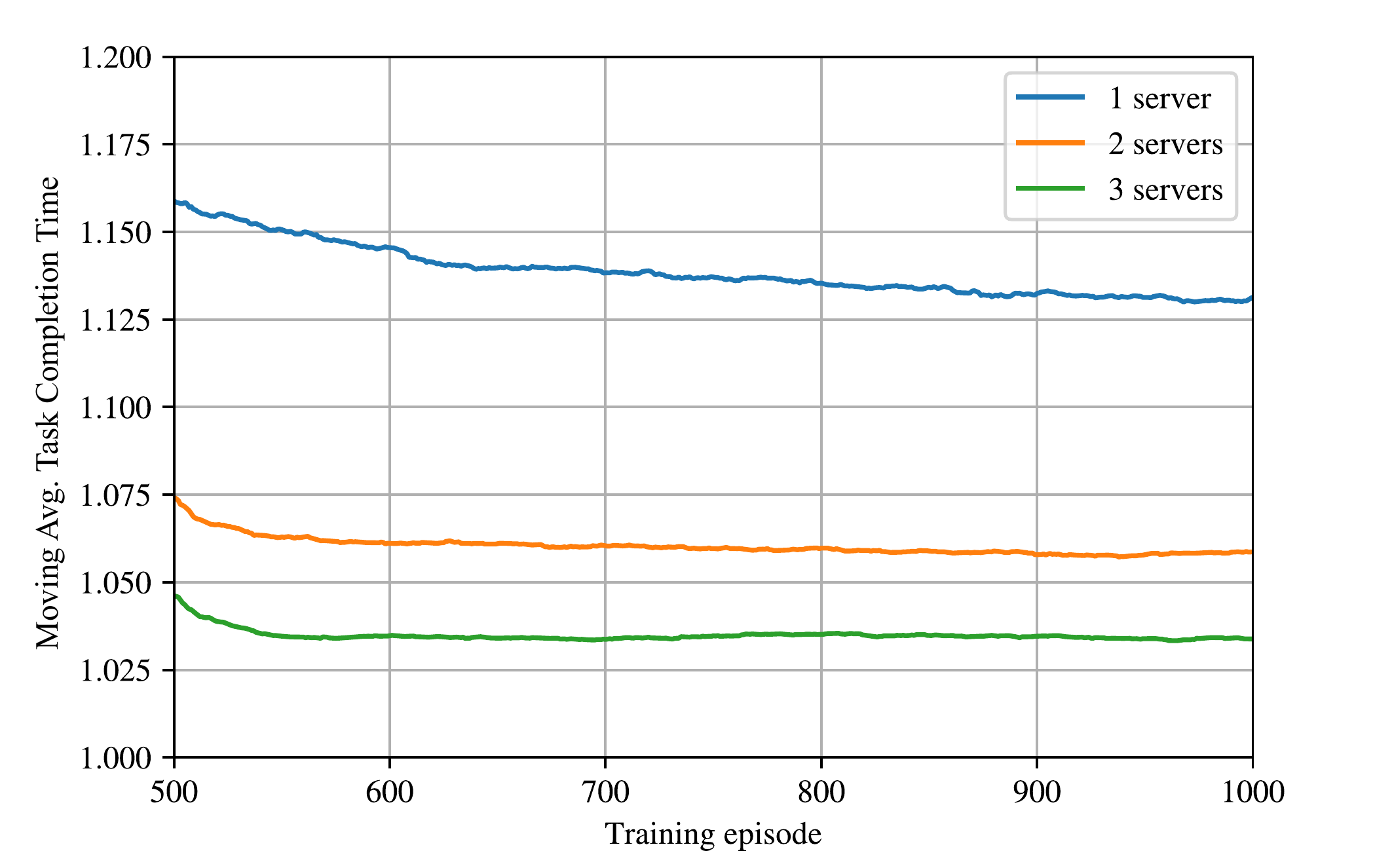}
}
\caption{Impact of the number of MEC servers on (a) system lifetime, and (b) task completion time (in terms of number of time intervals) for a system with 5 users.}
\label{fig:servers_result}
\end{figure*}

The plots in Figure~\ref{fig:servers_result} show the impact of the number of servers on the system performance in terms of lifetime and task completion time for a system with 5 users. 
As the plots show, the training process converges after around 1000 episodes. Moreover, our proposed approach confirms the fact that densifying the network with more MEC servers allows superior load balancing among them, hence improving the overall system performance.

\begin{figure*}[ht]
\centering
\vspace{-.5cm}
\subfloat[]{
  \includegraphics[width=0.4\textwidth]{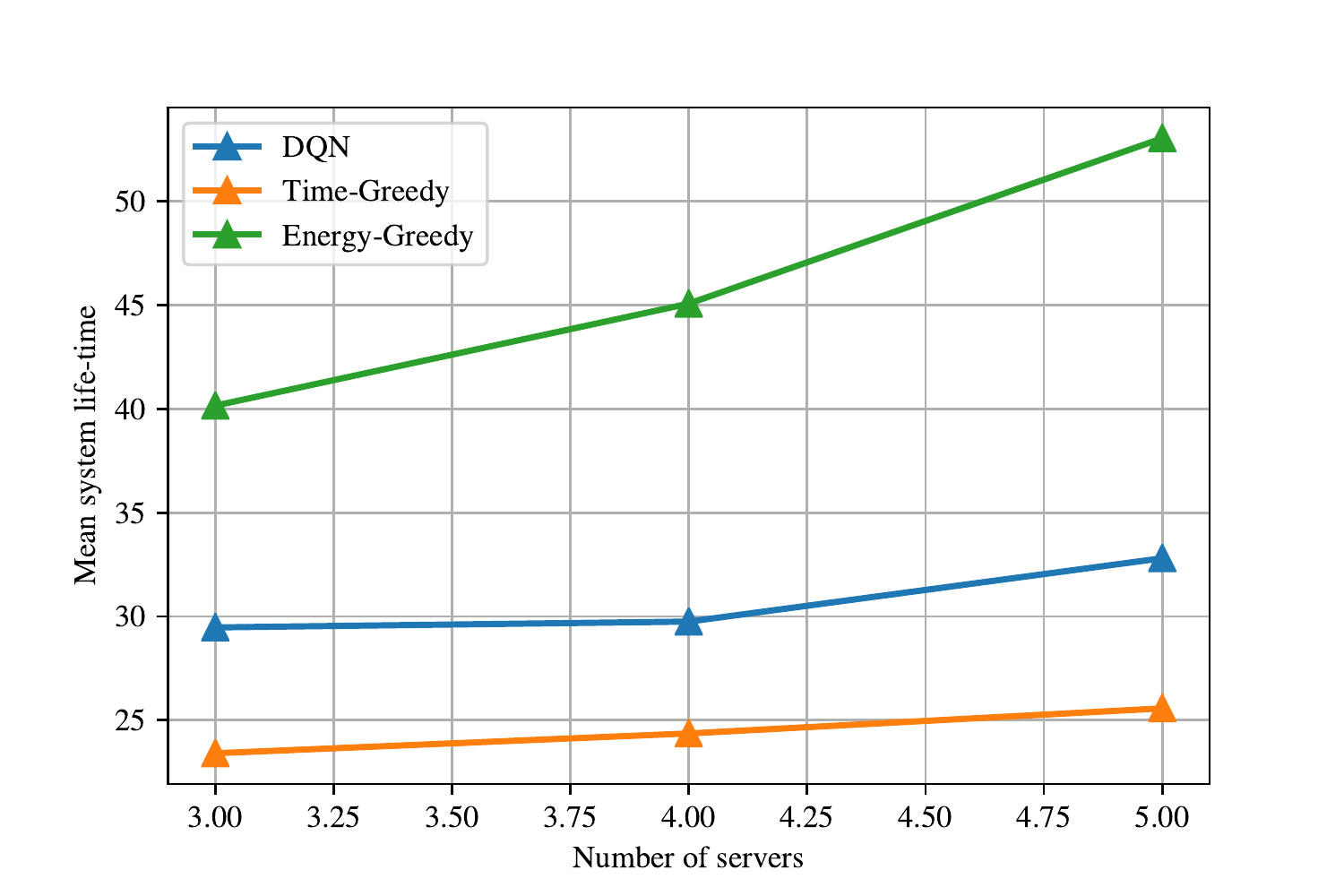}
}
\subfloat[]{
  \includegraphics[width=0.4\textwidth]{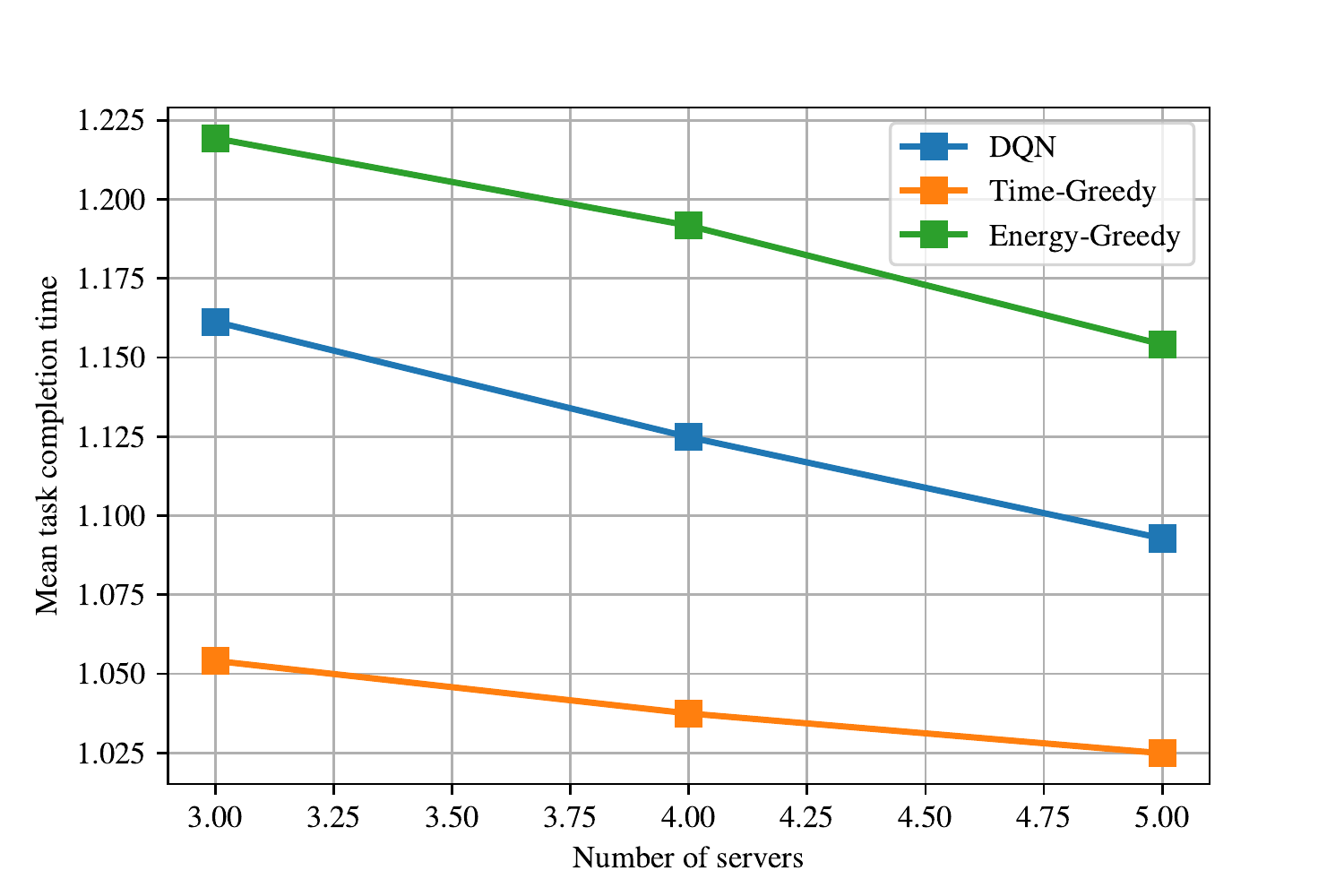}
}
\caption{(a) Lifetime comparison with greedy agents (b) Completion time comparison with greedy agents. The y-axis unit in both plots is in terms of number of time intervals.}
\label{fig:greedy_result}
\end{figure*}




In order to investigate the performance of our framework after training is complete, we define the following two greedy baseline agents:
\begin{itemize}
\item \textbf{Time-Greedy Agent:} This agent aims to minimize the task completion time by selecting the user with the largest average queue waiting time at each time interval. 
\item \textbf{Energy-Greedy Agent:} This agent is used to enhance the lifetime of the system by selecting the user with the lowest energy level at each time interval. 
\end{itemize}

In Figure \ref{fig:greedy_result}, we fix the network size to have 3 servers and 5 users, and compare the performance of our proposed DRL-based scheme with Time-Greedy and Energy-Greedy approaches. As the results show, our approach achieves a better trade-off between the mean task computation time and system lifetime compared to the aforementioned greedy agents. 


\section{Concluding Remarks}\label{sec:conc}
In this paper, we considered the problem of computation offloading in a mobile edge computing (MEC) architecture, where multiple energy-constrained users compete to offload their computational tasks to multiple servers. We developed a deep reinforcement learning framework
in which each server is equipped with a deep Q-network agent to select the best user for offloading at each time interval. Numerical results demonstrated the superiority of our approach over baseline algorithms in terms of the trade-off between task computation time and system lifetime.

\bibliographystyle{IEEEtran}
\bibliography{sample-bibliography}

\end{document}